# Stretched Exponential Relaxation of Glasses: Origin of the Mixed Alkali Effect

*Yingtian Yu[1], John C. Mauro[2], Mathieu Bauchy[1]*
[1] Physics of AmoRphous and Inorganic Solids Laboratory (PARISlab), Department of Civil and Environmental Engineering, University of California, Los Angeles, CA, USA
[2] Science and Technology Division, Corning Incorporated, Corning, NY, USA

## INTRODUCTION

As non-equilibrium materials, glasses continuously relax toward the supercooled liquid meta-stable equilibrium state [1, 2]. Although the myth of flowing glasses in cathedrals suggests otherwise [3], the dramatic increase of glass viscosity with decreasing temperature renders relaxation effectively "frozen" at ambient temperature. However, specific glass compositions can surprisingly deform over time, even at low temperature. This phenomenon is known as the thermometer effect [1, 4] and is usually attributed to the mixed alkali effect (MAE), which is observed in oxide glasses comprising at least two alkali oxides, $AO_2$ and $BO_2$, and manifests itself by a nonlinear evolution of the properties with respect to the fraction $A/(A + B)$.

Despite its practical importance [5], the nature of relaxation in the glassy state remains poorly understood. To date, no clear atomistic mechanism of structural and stress relaxation is available, which seriously limits our ability to predict and control their behavior [6]. This gap of knowledge becomes even more problematic as (1) the need for large screens with higher resolution (smaller pixel size) results in lower tolerances for relaxation, and (2) novel LCD fabrication processes (e.g., p-Si applications) require the use of higher processing temperatures, further enhancing relaxation [5]. Addressing these Grand Challenges [7] requires a better understanding of the fundamentals of glass relaxation, which is the goal of the present thesis project.

## ACCELERATED RELAXATION TECHNIQUE

Here, to investigate the MAE in glass relaxation, we simulated using molecular dynamics (MD) a series of $(K_2O)_x(Na_2O)_{16-x}(SiO_2)_{84}$ (mol %) mixed alkali silicate glasses, made of 2991 atoms, with varying $x$. All MD simulations were performed using the well-established Teter potential [8–10] with an integration timestep of 1 fs. Coulomb interactions were evaluated by the Ewald summation method with a cutoff of 12 Å. The short-range interaction cutoff was chosen as 8.0 Å. Liquids were first generated by placing the atoms randomly in the simulation box. The liquids were then equilibrated at 5000 K in the *NPT* ensemble for 1 ns, at zero pressure, to assure the loss of the memory of the initial configuration. Glasses were formed by linear cooling of the liquids from 5000 to 0 K with a cooling rate of 1 K/ps in the *NPT* ensemble at zero pressure.

To simulate the long-term relaxation of these glasses, we relied on a new accelerated simulation technique that we recently developed to understand the origin of room-temperature relaxation in Corning® Gorilla® Glass [1, 4]. In that method, the simulated glass is subjected to small, cyclic perturbations of volumetric stress. This method mimics the relaxation observed in granular materials subjected to vibrations [11], wherein small vibrations tend to densify the material (artificial aging), whereas large vibrations randomize the grain arrangements (rejuvenation). Similar ideas, relying on the energy landscape approach [12], have been applied to non-crystalline solids, based on the fact that small stresses deform the energy landscape locally explored by the atoms. This can result in the removal of some energy barriers that exist at zero stress, thus allowing the system to jump over the barriers to relax to lower energy states. More details about this method can be found in Ref. [1].

As expected, the stress perturbations allow all glasses to relax towards lower energy states [1]. As shown in Fig. 1, all glasses also show a gradual compaction in volume upon relaxation. Remarkably, the volume relaxation observed herein follows a stretched exponential decay function that is similar to that observed experimentally [4]. Further, we observe that the mixed $(K_2O)_8(Na_2O)_8(SiO_2)_{84}$ glass (denoted Na+K thereafter) features a larger densification than the binary sodium and potassium silicate glasses (denoted Na and K thereafter). This is a clear demonstration that the thermometer effect is indeed a manifestation of the MAE.

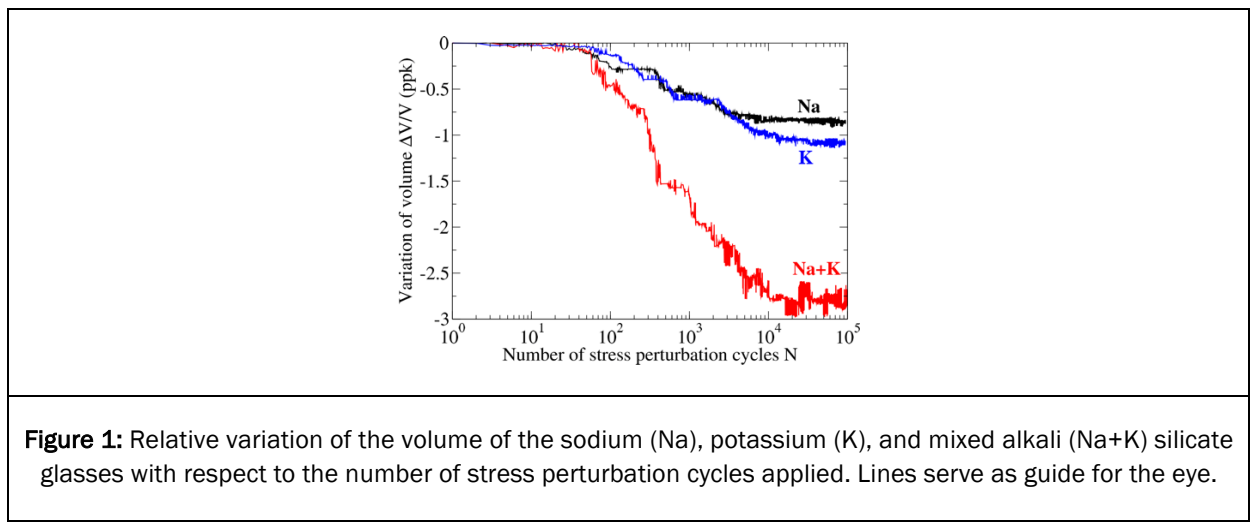

**Figure 1:** Relative variation of the volume of the sodium (Na), potassium (K), and mixed alkali (Na+K) silicate glasses with respect to the number of stress perturbation cycles applied. Lines serve as guide for the eye.

## ORIGIN OF THE MIXED ALKALI EFFECT IN RELAXATION

We now investigate the origin of the MAE in the context of relaxation. The stretched exponential nature of glass relaxation can be predicted by the Phillips' diffusion-trap model, wherein "excitations" in the glass diffuse toward randomly distributed "traps" [13]. However, this model remains largely axiomatic. Here, we propose that the excitations introduced within the diffusion-trap model correspond to locally unstable atomic units.

To assess this hypothesis, we first computed the coordination number (CN) of all atomic species. As shown in Fig. 2a, the CN of Na decreases upon the addition of K, whereas that of K increases upon the addition of Na, which can be attributed to a mismatch between the alkali atoms of the rest of the silicate network as one moves away from the binary composition. This miscoordinated state results in the formation of local stresses inside the atomic network, which was assessed by computing the local stress applied to each atom [14]. As shown in Fig. 2b, the average stress experienced by Na atoms increases upon the addition of K, whereas that experienced by K atoms decreases upon the addition of Na. This can be understood as follows. Over-coordinated K atoms present an excess of O atoms in their first coordination shell. Due to mutual repulsion, O atoms tend to separate from each other, which, in turn, tends to stretch the K–O bonds. On the other hand, under-coordinated Na atoms show a deficit of O atoms, which, in turn, are more attracted by the central cation. This results in a compression of Na–O bonds.

The mechanism of glass relaxation can then be understood as follows. Miscoordinated species act as local instabilities (or "excitations" following Phillips terminology). These excitations diffuse via local deformations of the atomic network, until an atomic arrangement that is locally under compression meets one that is under tension. At this point, both excitations are annihilated (or reach a "trap"), thereby relieving the initial internal stress stored in the network. The driving force for relaxation corresponds to the difference between the total cumulative stress experienced by Na and K atoms, which arises from the balance between two competitive behaviors. (1) The absolute stress per atom experienced by Na and K species increases upon the addition of K and Na, respectively. (2) In contrast, the numbers of Na and K atoms present in the network decreases upon their replacement by K and Na atoms, respectively. Altogether, as shown in Fig. 2c, the total cumulative stress experienced by Na and K atoms reaches a maximum when the number of Na equals that of K. This behavior provides an intuitive atomistic origin of the excessive volumic relaxation of glasses comprising mixed alkali atoms (i.e., thermometer effect).

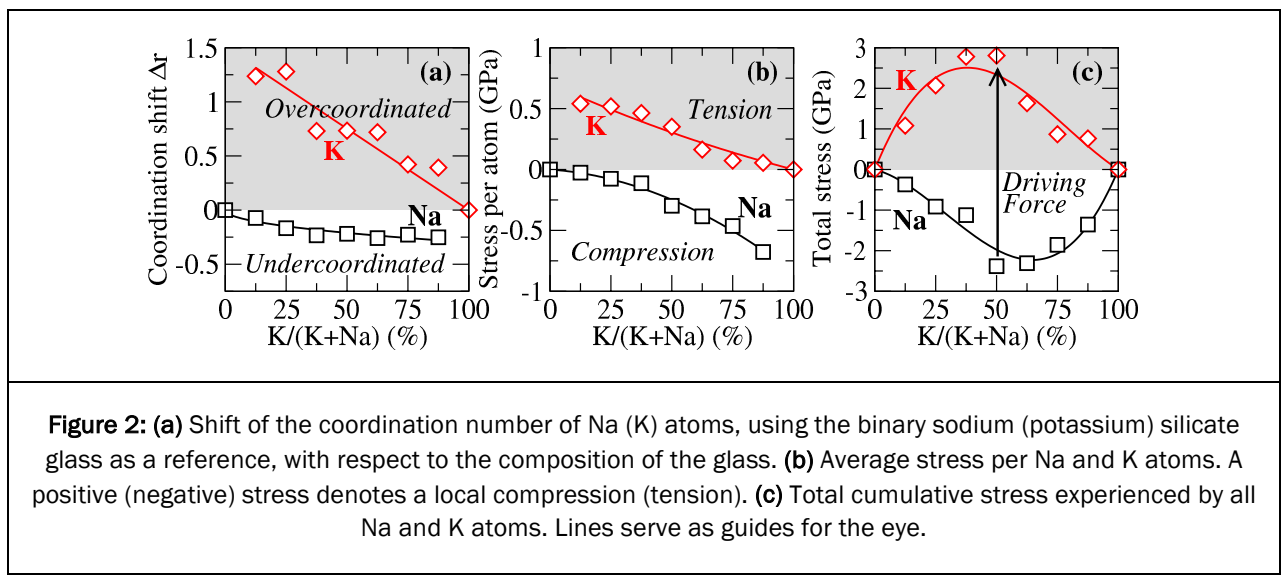

**Figure 2: (a)** Shift of the coordination number of Na (K) atoms, using the binary sodium (potassium) silicate glass as a reference, with respect to the composition of the glass. **(b)** Average stress per Na and K atoms. A positive (negative) stress denotes a local compression (tension). **(c)** Total cumulative stress experienced by all Na and K atoms. Lines serve as guides for the eye.

## FUTURE WORK

In future work, we plan to extent our analysis to mixed alkaline earth silicate glasses to assess the generality of the present results. Besides relaxation, we also plan to investigate the effect of the local atomic instabilities evidenced herein on the mechanical properties of glasses.